\documentclass[journal]{IEEEtran}

\usepackage{multirow,rotating,textcomp,booktabs}
\usepackage[ruled,vlined]{algorithm2e}
\usepackage{multirow,textcomp,booktabs}
\usepackage{mathrsfs}
\usepackage{graphics}
\usepackage{epsfig}
\usepackage{cite}
\usepackage{amsmath}
\usepackage{amssymb}
\usepackage{amsfonts}
\usepackage{color,colortbl}
 \usepackage{mathabx}
\usepackage{threeparttable}
\usepackage{accents}
\usepackage{hhline}
\usepackage{calc}
\usepackage{subfigure}
\usepackage{stfloats}

\usepackage{midfloat}
\usepackage{mathrsfs}
\usepackage{mdwtab}
\usepackage{array}
\usepackage{soul}
\usepackage{accents}

\newcommand{\mat}[1]{\boldsymbol{#1}}

\newcommand{\SubText}[2]{\ensuremath{{#1}_\textnormal{#2}}}
\newcommand{\ParComp}[2]{\parbox{#1}{\centering #2}}

\begin{document}
\title{An Algebraic Approach for the Stability Analysis of BLDC Motor Controllers}

\author{Julio C. G. ~Pimentel,~\IEEEmembership{Senior Member,~IEEE,}
        Emad ~Gad,~\IEEEmembership{Senior Member,~IEEE,}

\thanks{J. C. G. Pimentel (e-mail: pimentel@ieee.org) is with }% <-this % stops a space
\thanks{E. Gad is with SITE, University of Ottawa, Ottawa, ON, CA,
  e-mail: egad@uottawa.ca.}% <-this % stops a space
}


\markboth{Paper Submitted to IEEE ACCESS}%
{Shell \MakeLowercase{\textit{Pimentel et al.}}: Bare Demo of IEEEtran.cls for Journals}



\maketitle

%############################################################################
\begin{abstract}
% This paper discusses the stability issues of PI-based BLDCM
% (Brushless DC Motor) speed controllers under the presence of
% strong time-delay in the controller loop. Understanding of time-delay
% effect is of paramount importance to increase performance, quality and
% productivity in important applications such as electric differential
% for electric vehicles and high speed spindles in milling and high
% accuracy tools among others. Unfortunately, the effect of time-delays
% in the asymptotic stability of speed controllers has not been well
% studied in the existing literature.

This paper presents an algebraic technique to compute the
maximum time-delay that can be accepted in the control loop of a
Brushless DC Motor (BLDCM) speed  controller before the closed loop
response becomes unstable. Using a recently
proposed time-delay stability analysis methodology, we derive accurate 
stability conditions for the BLDCM speed controller. The results of
applying the new method show
that tuning the PI controller for 
very fast response in the order of magnitude of the BLDCM mechanical
time constant cause the time-delay to significantly affect the system
stability. 
\end{abstract}


% \fi
\IEEEpeerreviewmaketitle

%############################################################################
\section{Introduction}
\label{sec:introduction}

\IEEEPARstart{I}{n} 
the last decade, the brushless direct current motors (BLDCM) became widely used in a
variety of applications due to its robust mechanical topology and
simplicity of control, higher speed of operation, higher
torque for the same power density and lower manufacturing cost
compared to existing frequency controlled AC drives and vector
controlled permanent magnet synchronous motors (PMSM). They have also
become widely used in low power and high speed applications creating
a need for efficient and low cost controllers
\cite{xia2012:_bldc_motor_and_control}
\cite{Krishman09:_bldc_motor_and_control}. 
BLDC motors are also used in energy related applications such as 
hybrid vehicles integrated starter-generator, fuel pumps and electric
differential \cite{Chun14:_bldc_motor_control}
\cite{Collins12:_bldc_motor_control} \cite{Gougani12:_bldc_motor_control},
consumer appliances, computer numerical control, drilling 
tools, small hydro and wind energy generation, and flywheel
energy-storage systems \cite{vanisri11:_torque_ripple_minimization} 
\cite{Kolondzovski10:_power_limits_of_high_speed_machines}.

Nonetheless, the industrial potentials of the BLDCM pose new challenges
for the close-loop control design that were not seen with the
controllers of classical motors. One such challenge arises from the
small mechanical time constant  ($\tau_{\textnormal{mech}}$), which
typically approaches the order of few milliseconds. With such a
small time constant, the total time delay in the controller
($\tau_{\mathrm{total}}$) induced by the various modules in the
closed-loop becomes a dominant player in determining the stability of
the controller. 

The question of whether a closed-loop controller with a particular
delay value is stable or not is easily answered through the
classical graphical methods on Bode or Nyquist plots
\cite{franklin98:_digital_control}
\cite{ogata95:_discrete_time}. However, in the context of the BLDCM, the more fitting question to ask is: \textit{how much total delay can be tolerated in the closed-loop before the system exhibits unstable behaviour}? The lack of a satisfactory, accurate and simple method to answer that question typically forces the motion control designer to use conservative tuning scheme, a practice that often times comes at the expense of slowing down the response of the set-point tracking or the load-disturbance rejection.

The purpose of this paper is to present a simple algebraic approach to answer the above question. More precisely, the proposed method enables computing the maximum delay $\tau_{\mathrm{max}}$ that can be allowed in the control loop while maintaining the desired margins of stability in the system.

The immediate benefit gained from the new method of the proposed method is that it provides, so to speak, a new lens through which commonly used PI- controller tuning methodologies can be viewed and assessed. Indeed, as will be shown in this paper, the proposed approach offers new insights in the famous tuning methodologies that remained hitherto unknown. The long-term benefits of the proposed method is that it opens the door to new automatic tuning strategies that take into account the actual values of $\tau_{\mathrm{total}}$ and $\tau_{\mathrm{max}}$ in tuning the controller parameters. 

The proposed method is based on a recent approach, initially proposed in \cite{Olgac02:_rtds_stability}, and later extended in \cite{Olgac04:_rtds_stability}, to derive the stability condition of a linear time invariant retarded time-delay system (LTI-RTDS).  The proposed technique constructs an analytical model for the controller of the BLDCM set-point tracking and load-disturbance rejection transfer functions that takes into account the various sources of delay in the control loop. It then adapts the method of \cite{Olgac04:_rtds_stability} to estimate the maximum delay, $\tau_{\max}$ that can be tolerated in the loop before the system becomes unstable. Subsequently, $\tau_{\max}$ arising from commonly-used PI controller tuning methodologies is computed, and used to shed the light on the performance of the tuning methodology. Future works will use the proposed method to derive a systematic procedure to tune the controller parameters to achieve a more optimized performance.

The rest of the paper is organized as follows. Section \ref{sec:methodology}
presents a quick summary of the time-delay analysis method
used. Section \ref{sec:delay-based-modell} briefly develops
the BLDCM state space model and identifies various sources of delay
affecting the controller stability and presents the development of the
BLDCM models with time-delays. Section \ref{sec:effect-delay-stabl}
derives the stability conditions and analyzes the effect of the delay
on the stability of the closed-loop control system. In section
\ref{sec:effect-controller-stabl} we analyze the effect on the
stability condition of varying the controller parameters. Finally, section
\ref{sec:experimental_results} presents the test bench built to
validate the results including simulated and measured results.

%############################################################################
\section{Background, Motivation and Problem formulation}
\label{sec:backgr-motiv-probll}
This section sets the stage for the problem scope addressed by the
work presented in this paper. It also lays out the motivation the
relavant mathematical problems formulation.

\subsection{Problem Scope}
\label{sec:problem-scope}
\begin{figure}[htbp!]
  \centering
  \includegraphics[width=0.45\textwidth]{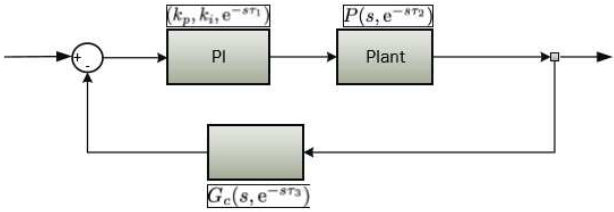}
  \caption{PI-based control with several delay sources.}
  \label{fig:Feedback-system}
\end{figure}

Fig. \ref{fig:Feedback-system} depicts a representation of the scope
of the problem addressed by the approach presented in this work. The
plane in the block diagram of Fig. \ref{fig:Feedback-system}
represents the BLDC moto and the PI block represents the
Proportional-Integrator controller module. The transfer function of
the plant may include certain elements that cause a pure delay which
is taken into account by expressing the transfer function as a
function of two variables $s$ and $\mathrm{e}^{-s\tau_2}$. Likewise
the PI controller may also include a delay $\mathrm{e}^{-s\tau_1}$ in
addition to the classical proportional and integrator constants $k_p$
and $k_i$, respectively. The feedback path is also a delay-dependent
transfer function $G_c(s,\mathrm{e}^{-s\tau_3})$

The delays in the above components are assumed to be characterized by
uncertainty or may alternatively be regarded as shifting with time in
an unspecified way. Those delays arise not from design decisions, but only as 
second-order effect from the system hardware or wiring.

It is also assumed that absent those sources of delays,
$\tau_1=\tau_2=\tau_3=0$, the close-loop control system is stable. On
the other hand, the presence of the delays, i.e. $\tau_i>0$ may, or
may not, render the system unstable. However, it is not the main
concern in this paper to determine whether the system is stable for
particular values of the delays. Rather, the main concerns in this
work can be summed up by the following questions:
\begin{enumerate}
\item Is there a limit, efficiently computable, for those delays beyond
  which, the system becomes unstable.
  \item If that limit is found to exist, then how is this limit
    compared with the actual delays in the circuit. 
    \item Denoting the difference
    between the limit and actual delays in the system by the
    so-called \textit{delay margin}, what is the impact of $k_p$ and
    $k_i$ on the delay margin.
\end{enumerate}

\subsection{Motivations}
\label{sec:motivations}
In many systems depicted by the block diagram of
Fig. \ref{fig:Feedback-system}, the actual delays are
negligible in comparison with the time constants of the mechanical
dynamics of the system. Indeed, these situations  do not warrant the
investigations by the methods proposed in this work. However, in the
case of the BLDC motors, the mechanical time sontants is sufficiently
small that the delays in the closed-loop control system is a sizeable
portion from it. Under those conditions, the delay margin becomes an
important factor to take into account in designing or tuning the PI
parameters $k_p$ and $k_i$. For example, a desirable PI design would
be one that maximizes the delay margin to guard against potential
delay-caused instability but not at the expense of slowing down the
system extensively.

The formost goal of this work is to develop a
simple algebraic method that maps the various design parameters,
including the parameters $k_p$ and $k_i$, to the delay margin of the
system. The proposed method will used to take a new look at the
existing PI design tuning methodologies and show through experimental
results their impact on the system performance.

\subsection {Mathematical Problem Formulation}
\label{sec:probl-form-sect}

The mathematical model that represets the close-loop BLDC motor
problem is cast as a system of linear time-invariant retarded
time-delay system (LTI-RTDS) that takes the following form

\begin{equation}
  \label{eq:1}
  \frac{\textnormal{d}\mat{x}(t)}{\textnormal{d}t} = \mat{A}_0\mat{x}(t) + \mat{A}_1\mat{x}(t-\tau) +\mat{B}\mat{u}(t)
\end{equation}
The derivation of the above formulation for the underlying system will
detailed in Section \ref{sec:delay-based-modell}. The parameters in \eqref{eq:1} are defined as follows. $\mat{x}(t)\in\mathbb{R}^n$ is the state-space vector,
$\mat{A}_0$ and $\mat{A}_1\in\mathbb{R}^{n\times n}$ are
real coefficient matrices with ranks $n,N$, respectively, ($N \le n$),
$\tau \in \mathbb{R}^+$ is a parameter that represents the delay in
the system, $\mat{B}\in\mathbb{R}^{n\times p}$ is the input matrix and
$\mat{u}(t)\in\mathbb{R}^p$ is called the input (or control) vector.

The system is assumed to be stable without delay,
i.e., if $\tau=0$. One must also note here that the delay parameter $\tau$ in \eqref{eq:1} is the actual delay, that
is, it is a delay introduced (inadvertently) by the various modules in the closed-loop
control system. The first objective in the paper will be to develop a
method to compute the so-called delay margin, which is the difference
between the system actual delay, $\tau$, and the maximal delay limit,
denoted $\SubText{\tau}{max}$, beyond which the system becomes
unstable. The underlying assumption here is that the increase in the value
of $\tau$ pushes the system away from stability and closer to
instability. Thus, the system is unstable if $\tau >
\SubText{\tau}{max}$, and stable if $\tau <
\SubText{\tau}{max}$. Furthermore, the system is stable independent of
the delay if $\SubText{\tau}{max} = \infty$. $\SubText{\tau}{max}$ considered above is a function of the system
parameters, or more precisely, the entries of the
matrices $\mat{A}_0$ and $\mat{A}_1$, whereas $\tau$ is independent of
those parameters. Therefore, the task of computing the delay margin in
the system can be viewed as the task of $\SubText{\tau}{max}$. Section
\ref{sec:methodology} presents the method to compute $\SubText{\tau}{max}$.

\subsection{Related Work}
\label{sec:related-work}
Investigating the effect of the delay on the stability of the retarded
system dates back to several decade ago.
Classical methods based on Bode or Nyquist plots are by far the
preferred methods used to analyze the effect of the time-delay on
the stability of digital speed controllers with loop time-delays
\cite{franklin98:_digital_control} \cite{ogata95:_discrete_time}.
That is probably the case because they are easy to use and provide some insight if
one seeks to investigate the stability of the system for a particular
value of delay $\tau$. However,
they are rather inconvenient if we either need to analyze the effect of a
range of delay values or find the maximum loop time-delay
beyond which the system becomes unstable.

During the last decade, other methods have been proposed to address
this issue. Despite the fact that some accurate results have been
reported, those methods are often computationally expensive for
high-order systems because  they map the original time-delay
analysis to solving an equivalent LMI (Linear matrix Inequality)
problem
\cite{Guoping_Lu_PR_delay_2000,Delay_effect_stability,Gu03b:_time_delay_systems,Niculescu04:_time_delay_system,Atay10:_time_delay_systems,Delice12:_time_delay_systems,Ghaoui00:_time_delay_systems}.

\section{Computing Maximal Stable Delay in LTI-RTDS, $\tau_{\max}$}
\label{sec:methodology}
The objective in this section is to consider a system described by the
LTI-RTDS described by \eqref{eq:1} and ask the question: \textit{what maximal value for the delay $\tau$ will turn
the system to unstable system assuming that the system at $\tau=0$ to
be intrinsically stable}? The procedure described in this section has been
presented in \cite{Olgac02:_rtds_stability}.  The theoretical background to this
procedure has been presented in \cite{Olgac02:_rtds_stability}
\cite{Olgac04:_rtds_stability}, and a comparative study with other
methodolgies has been presented in \cite{Sipahi05:_rtds_stability}. A
simplified and systematic application of this procedure is presented
in the next subsection.  We should also note that,
in line of the assumption made about the system in \eqref{eq:1}, this method assumes that the delay-free system (i.e., $\tau=0$) is stable by construction.

In general, the system is asymptotically stable for a given $\tau$ if, and only if, the roots of the characteristic polynomial obtained from 
\begin{equation}
  \label{eq:det-ch-poly}
  \textnormal{CE}(s,\tau) := \det\left(s\mat{I} - \mat{A}_0 - \mat{A}_1e^{-s\tau}\right) 
\end{equation}
or, alternatively, defined by
\begin{equation}
  \label{eq:det-ch-poly-1}
  \textnormal{CE}(s,\tau) = \sum_{k=0}^np_k(s)e^{-sk\tau} 
\end{equation}
are all in the left-half plan of the complex $s$ plan. $p_k(s)$ in the above equation is polynomial in $s$ of degree $n-k$ with real coefficients.

The transcendental nature of $\textnormal{CE}(s,\tau)$ produces an infinite number of roots, thereby making the task of analyzing the stability for given $\tau$ very complex, and finding $\tau_{\max}$ even more cumbersome.

\subsection{Description of the Basic Procedure}
\label{sec:descr-basic-proc}
In order to facilitate the description of finding $\tau_{\max}$ of a general LTI-RTDS, the following presentation will consider its application to an example LTI-RTDS given by
\begin{equation}
  \label{eq:An-exmaple}
  \mat{A}_0   = \left[ \begin{smallmatrix} -2.0  &  0.0 \\  0.0  & -0.9 \\ \end{smallmatrix} \right], \mat{A_1} = \left[ \begin{smallmatrix} -1.0  &  0.0 \\ -1.0  & -1.0 \\ \end{smallmatrix} \right]
\end{equation}
It is worth noting that for this example, the exact value of $\tau_{\max}$
is known \textit{a priori} using an analytical argument as has been
shown in \cite{Gu03b:_time_delay_systems}. This fact will be used to
validate the result obtained from the procedure below with the exact
solution. The procedure can be described as sequence of 6 steps
summarized next.

\begin{itemize}
\item \textbf{Step 1.} Use the the Rekasius mapping to map $e^{-s\tau}$ as follows
  \begin{equation}
    \label{eq:mapping-1}
    e^{-s\tau} = \frac{1-sT}{1+sT}, T\in\mathbb{R}
  \end{equation}
where $T$ is related to $\tau$ through the following relation
\begin{equation}
  \label{eq:mapping-2}
  \tau = \frac{2}{\omega} \left(\tan^{-1}\left(wT\right)\mp l\pi\right), \quad l = 0,1,2,\cdots
\end{equation}
The above mapping transforms $\textnormal{CE}(s,\tau)$ of
\eqref{eq:det-ch-poly-1} into a polynomial of degree $2n$ in $s$,
whose coefficients are polynomials in $T$,
\begin{equation}
  \label{eq:CE-transformed-T}
  \widebar{\textnormal{CE}}(s,T) = \sum_{j=0}^{2n} q_j(T) s^j
\end{equation}

It is crucial to stress the fact that the Rekasius mapping is exact
for $s=\j\omega$, $\omega\in\mathbb{R}$, in the sense that
$\widebar{\textnormal{CE}}(s,T) = \textnormal{CE}(s,\tau)$ $\forall s
= \j\omega$, where $\j=\sqrt{-1}$.

In the example, taken for demonstration $n=2$. This step would lead to the following polynomials
\begin{eqnarray*}
  \label{eq:Routh_hurwitz-array-application-1}
  q_{4}(T) &=& T^2\\
  q_{3}(T) & =& 0.9T^2 +2T\\
  q_2(T) & = &-0.1T^2 + 5.8T+1\\
  q_1(T) & = &1.6T+4.9\\
  q_0(T) & = &5.7\\
\end{eqnarray*}
\item \textbf{Step 2.} Form the Routh-Hurwitz array\cite{laughton02:_elect_engin_refer_book} for the $s$
  polynomial in \eqref{eq:CE-transformed-T} 

  \begin{equation*}
    \label{eq:Routh-Hurwitz-array}
    {\text{\small{
    \begin{array}[c]{c | c  c c c c}
      s^{2n}&q_{2n}(T) & q_{2n-2}(T)& q_{2n-4}(T) & \cdots & q_0(T)\\
      s^{2n-1}&     q_{2n-1}(T) & q_{2n-3}(T)& \cdots & q_1(T) & 0\\
      s^{2n-2}&  v_{1}^{(2n-2) } (T)& v_{2}^{(2n-2) }(T) &v_{3}^{(2n-2) }(T)& \cdots&0\\
      s^{2n-3}& v_{1}^{(2n-3) } (T)& v_{2}^{(2n-3) } (T)&v_{3}^{(2n-3) }(T) &\cdots&0\\
      \vdots& \vdots& &  &&\\
      s^2& v_1^{(2)}(T)&      v_2^{(2)}(T)& 0 & \cdots&0\\
      s^1&     v_1^{(1)}(T) & 0 &\cdots & \cdots&0
    \end{array}}}}
  \end{equation*}
The application of this step to the particular example considered above will result in
\begin{eqnarray*}
  \label{eq:Routh_hurwitz-array-application-2}
    \text{\scriptsize{$v_1^{(2)}(T)$}} &\text{\scriptsize{$=$}}& \text{\scriptsize{$\frac{-0.09T^4 + 3.42T^3 + 7.6T^2 + 2T}{0.9T^2 + 2T}$}}\\
    \text{\scriptsize{$v_2^{(2)}(T)$}} &\text{\scriptsize{$=$}}& \text{\scriptsize{$5.7$}} \\
    \text{\scriptsize{$v_1^{(1)}(T)$}} &\text{\scriptsize{$=$}}& \text{\scriptsize{$\frac{-0.144T^5 + 0.414T^4 + 8.4T^3 + 17.64T^2 + 9.8T}{-0.09T^4 + 3.42T^3 + 7.6T^2 + 2T}$}}
\end{eqnarray*}
\item \textbf{Step 3.} Compute the roots of $v_1^{(1)}(T) = 0$. This set of roots is referred to as $T_{cr}$, and for the current example are given by
  \begin{equation*}
    T_{cr }=  \{
        \begin{array}[c c c c c]{c c c c c}
          -4.67,& -2.22,& -1.46,& -1.0,& 10.0
        \end{array}
        \}
  \end{equation*}
\item \textbf{Step 4.} Compute $\omega_{cr}$ using $T_{cr}$ from
  \begin{equation}
    \label{eq:omega-cr}
    \omega_{cr} = \sqrt{{\frac{ v_2^{(2)}(T_{cr})} { v_1^{(2)}(T_{cr})}}}
  \end{equation}
Note that $T_{cr}$ (and consequently $\omega_{cr}$) depends only on the state-space matrices $\mat{A}_0$ and $\mat{A}_1$.
Defining $\omega_{cr}^{+} \subseteq \omega_{cr}$ as the subset of $\omega_{cr}$ with strictly positive values, and the corresponding $T_{cr}$ values as $T_{cr}^{+}$, use \eqref{eq:mapping-2} to compute
\begin{equation}
  \label{eq:tau-cr}
  \tau_{cr}^{+} = \frac{2}{\omega_{cr}} \left(\tan^{-1}\left(\omega_{cr}T_{cr}\right) \mp l\pi\right), \quad l=0,1,2,\cdots
\end{equation}

In the one-dimensional parameter space, $\tau$, the above set represents the boundaries of the stable and unstable regions of delay of the LTI-RTDS. 

\item \textbf{Step 5.} Compute the Root Tendency (RT) using
\begin{equation}
  \label{eq:rt_compute_equation}
  RT = \textnormal{sgn} \left[ \Im \left( \frac{\sum\limits_{k=0}^N a'_k e^{-s k 
          \tau}} {\sum\limits_{k=0}^N k a_k e^{-s k \tau}} \right) \right]
\end{equation}
where $s=\j\omega_{cr}^{+} $, $\tau = \tau_{cr}^{+}$, $\Im$ denotes the imaginary part and ``sgn'' is the sign ($\pm 1$).
$RT$ represents the root transition direction 
crossing the imaginary axis to the unstable
Right-Half Plan (RHP) ($RT=+1$) or to the stable Left-Half Plan (LHP)
($RT=-1$).

In the context of applying this procedure to the stability analysis of
the BLDCM speed controller, it is typically the case that the
delay-free system ($\tau=0$) is stable by design. The question of
finding the maximum delay of stable operation
$\tau_{\textnormal{max}}$ then becomes finding the minimum member of
the set $\tau_{cr}^{+}$ such that $RT=+1$. This is the basis for the final step.

\item \textbf{Step 6.} Compute $\tau_{\textnormal{max}}$ using 
\begin{equation}
  \label{eq:2}
  \tau_{\textnormal{max}} = \left\{ \begin{array}{ccc}
                             \min{\tau_{cr}^{+}} & \textnormal{if} & \exists{RT=+1}\\
                             \infty              & \textnormal{otherwise} & 
                             \end{array} \right.
\end{equation}

In the sense of \eqref{eq:2}, $\tau_{\textnormal{max}} = \infty$
implies that the LTI-RTDS is stable independent of the delay. The application
of the last three steps to the example test case, yields 
  \begin{equation}
    \label{eq:omega-cr-application}
    \omega_{cr}^{+} = 0.4359, \quad RT=+1, \quad \tau_{\max} =6.1726 
  \end{equation}

\end{itemize}

%############################################################################
\section{Delay-Based Modelling of the BLDCM Control Loop}
\label{sec:delay-based-modell}
This section turns the focus on the BLDC motor control loop. The ultimate
objective in this section is to show how the closed-loop control of
the BLDCM is properly cast as an LTI-RTDS of the form in \eqref{eq:1}.

To this end, the section first presents a brief background for the modeling of the BLDCM and the digital control loop. Section \ref{sec:bldcm_state_space_model} first describes a delay-free state-space model of the BLDCM motor. Section \ref{sec:speed_controller_with_delay} uses a closed-loop speed controller to derive a delay-based model for the transfer function of the set-point tracking and load disturbance rejection.

\subsection{BLDCM State Space Model}
\label{sec:bldcm_state_space_model}

Modeling of BLDC motor has been well studied in the literature, e.g. \cite{xia2012:_bldc_motor_and_control}
\cite{Godoy02} \cite{Yeong07} \cite{Kim08}. Assuming that the BLDCM of
Fig. \ref{fig:bldcm_motor_electric} is symmetric in all three phases
and that there is no change in rotor reluctance with angle because of
a non-salient rotor, its electrical circuitry model can be written as:
\begin{figure}[hbtp]
  \centering
  \includegraphics[width=0.45\textwidth]{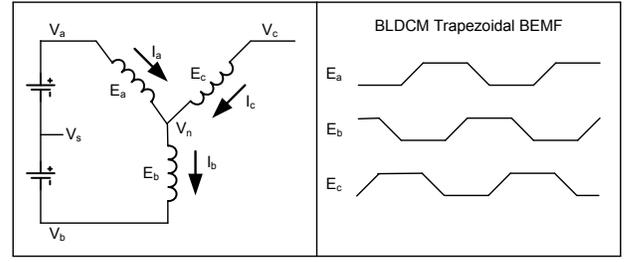}
  \caption{BLDC Motor Electric Circuitry.}
  \label{fig:bldcm_motor_electric}
\end{figure}

\begin{IEEEeqnarray}{rcl}
  \label{eq:bldc_elect_model}
  \left[ \begin{smallmatrix} v_a \\ v_b \\ v_c \end{smallmatrix} \right] =
  \left( \left[ \begin{smallmatrix} R_s & 0 & 0 \\ 0 & R_s & 0 \\ 0 & 0 & R_s \end{smallmatrix} \right] + 
  \frac{d}{dt} \left[ \begin{smallmatrix} L_s & 0 & 0 \\ 0 & L_s & 0 \\ 0 & 0 & L_s \end{smallmatrix} \right] \right)
  \left[ \begin{smallmatrix} i_a \\ i_b \\ i_c \end{smallmatrix} \right] +
  \left[ \begin{smallmatrix} e_a \\ e_b \\ e_c \end{smallmatrix} \right]
\end{IEEEeqnarray}
where $v_a, v_b, v_c, i_a, i_b$ and $i_c$ are the motor phase voltage and
currents respectively, $L$ and $M$ are the
winding self and mutual inductance, $L_s = L - M$ and $e_a, e_b$ and
$e_c$ are the induced BEMF voltages. In a PMM, the BEMF is a function
of the rotor position and can be written as $e(\theta) =  \lambda
\omega_r f(\theta)$, where $\lambda$ represents the total flux linkage,
$\omega_r$ is the motor shaft rotational speed. For a BLDCM, $f(\theta)$ is a
trapezoidal function with peak values at $+1$ and $-1$. For the sake
of clarity, from now on we will omit the angle $\theta$ in the BEMF equation. 

The generated electromagnetic torque is given by equation
\eqref{eq:bldc_torque_equation}. If $J$ is the rotor moment of inertia,
$B_m$ is the viscous friction coefficient and $T_l$ is the load
torque, then the mechanical model can be written as in
\eqref{eq:bldc_speed_equation}.
\begin{equation}
  \label{eq:bldc_torque_equation}
  T_e = \frac{e_a i_a + e_b i_b + e_c i_c}{\omega_r} 
\end{equation}
\begin{equation}
  \label{eq:bldc_speed_equation}
  J \frac{d \omega_r}{dt} + B_m \omega_r = T_e - T_l 
\end{equation}
\begin{equation}
  \label{eq:bldc_position_equation}
  \frac{d \theta}{dt} = \omega_r 
\end{equation}

The state space model with $\mat{x}(t) = \begin{bmatrix}
  i_a ~i_b ~i_c ~\omega_r \end{bmatrix}^T$ and $\mat{u}(t) =
\begin{bmatrix} v_a ~v_b ~v_c ~T_l \end{bmatrix}^T$can be written as:
\begin{eqnarray}
  \label{eq:bldc_state_space_model}
  \frac{\textnormal{d}\mat{x}(t)}{\textnormal{d}t} & = & \mat{A} \mat{x}(t) + \mat{B}\mat{u}(t) \\
  \mat{y(t)} & = & \mat{C} \mat{x}(t) \IEEEnonumber
\end{eqnarray}
\begin{eqnarray}
  \mat{A} = \left[ \begin{smallmatrix}
      -R/L     & 0        & 0        & \lambda f_a(\theta)/L & 0 \\
      0        & -R/L     & 0        & \lambda f_b(\theta)/L & 0 \\
      0        & 0        & -R/L     & \lambda f_c(\theta)/L & 0 \\
      \lambda f_a(\theta)/J & \lambda f_b(\theta)/J & \lambda f_c(\theta)/J & -B_m/J
  \end{smallmatrix} \right] \IEEEnonumber
\end{eqnarray}
\begin{eqnarray}
  \mat{B} = \left[ \begin{smallmatrix}
      1/L & 0   & 0   & 0    \\
      0   & 1/L & 0   & 0    \\
      0   & 0   & 1/L & 0    \\
      0   & 0   & 0   & -1/J
  \end{smallmatrix} \right] \IEEEnonumber
\end{eqnarray}

Assuming the BLDC motor is phase-balanced and wye-connected then $i_a
+ i_b + i_c = 0$ and $v_s = \displaystyle\sum\limits_{i=a}^c v_i -
\displaystyle\sum\limits_{i=a}^c e_i$. Note that the motor can be
modeled by just two currents as the third current is dependent of the
other two. From the previous equations, we can derive the
BLDCM non linear state space model with state variables $i_a$, $i_b$
and $\omega_r$, given by:
\begin{eqnarray}
  \mat{A} = \left[ \begin{smallmatrix}
      -R/L     & 0        & \lambda/3L(2f_a-f_b-f_c) \\
      0        & -R/L     & \lambda/3L(2f_b-f_a-f_c) \\
      \lambda/2J(f_a-f_c) & \lambda/2J(f_b-f_c) & -B_m/J
  \end{smallmatrix} \right] \IEEEnonumber
\end{eqnarray}
\begin{eqnarray}
  \mat{B} = \frac{1}{3L} \left[ \begin{smallmatrix}
       2 & -1  & -1 & 0    \\
      -1 &  2  & -1 & 0    \\
       0 &  0  &  0 & -3L/J
  \end{smallmatrix} \right] \IEEEnonumber
\end{eqnarray}
where the state variable $\mat{x}(t)\in\mathbb{R}^{3}$ is given by
$\mat{x}(t) = \begin{bmatrix} i_a ~i_b ~\omega_r \end{bmatrix}^T$ and the
input vector $\mat{u}(t)\in\mathbb{R}^{4}$ is given by $\mat{u}(t) =
\begin{bmatrix} v_a ~v_b ~v_c ~T_l
\end{bmatrix}^T$. $\mat{A}\in\mathbb{R}^{3 \times 3}$,
$\mat{B}\in\mathbb{R}^{3 \times 4}$ and
$\mat{C}=\mat{I}\in\mathbb{R}^{3 \times 3}$ are the matrices describing the
dynamics of the BLDCM continuous-time model (CTM).
We can further simplify the model in \eqref{eq:bldc_state_space_model}
to make it easier to analyze the BLDCM dynamical behavior as a function of its
mechanical and electrical parameters. In a BLDCM, at any time there
are only two phases being driven while the third phase is
open. Assuming that $B_m \ll 0$ such that $B_m R \approx 0$ and $B_m L \approx
0$, and that phases $a$ and $b$ are driven by a voltage source
$v_a$ and $v_b$ respectively, then $i_c=0$ and $i_a=-i_b$. Therefore,
the model in \eqref{eq:bldc_state_space_model}, with $k_e=2\lambda$,
$\mat{x}(t) = \begin{bmatrix} i_a ~\omega_r \end{bmatrix}^T$,
$\mat{u}(t) = \begin{bmatrix} (v_a-v_b) ~T_l \end{bmatrix}^T$, can be rewritten as:
\begin{eqnarray}
  \label{eq:bldc_state_space_model_2}
  \frac{\textnormal{d}\mat{x}(t)}{\textnormal{d}t} & = & \mat{A}_{2 \times 2} \mat{x}(t) + \mat{B}_{2 \times 2}\mat{u}(t) \\
  \mat{y(t)} & = & \mat{C}_{2 \times 2} \mat{x}(t) \IEEEnonumber
\end{eqnarray}
\begin{eqnarray}
  \mat{A}_{2 \times 2} = \left[ \begin{smallmatrix}
      -R/L    & -k_e/L \\
      k_e/J   & -B_m/J
  \end{smallmatrix} \right] \IEEEnonumber
\end{eqnarray}
\begin{eqnarray}
  \mat{B}_{2 \times 2} = \left[ \begin{smallmatrix}
       1/L & 0    \\
       0   & -1/J
  \end{smallmatrix} \right] \IEEEnonumber
\end{eqnarray}

%############################################################################
\subsection{BLDCM Speed Controller With Loop Delays}
\label{sec:speed_controller_with_delay}
\begin{figure*}[hbtp]
  \centering
  \includegraphics[width=0.8\textwidth]{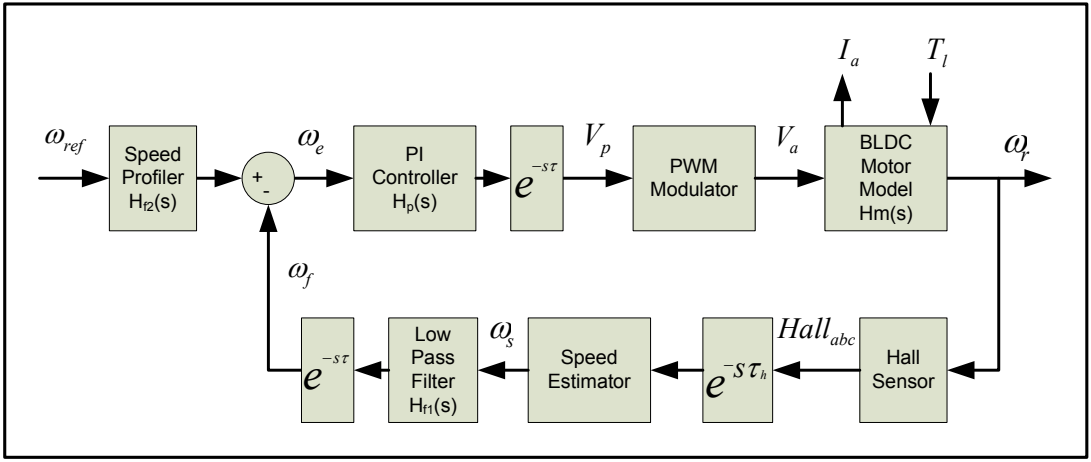}
  \caption{BLDC motor speed controller with sources of delay.}
  \label{fig:bldc_speed_controller_delay}
\end{figure*}

This section derives a closed-form for the BLDCM speed controller that includes the control loop delays. Figure \ref{fig:bldc_speed_controller_delay} presents the linearized
model of a BLDCM speed controller showing various sources of delay
in the control loop. As the figure shows, there are three sources of delay in the control loop. Those are
\begin{enumerate}
\item The delay introduced by the Hall sensor, $\tau_h$, can be
  estimated using the rotational speed $\omega_r$ in $rad/s$ as follows
  \begin{equation}
    \label{eq:Tau-Hall}
    \tau_h = \frac{2 \pi }{ 6 \omega_r}
  \end{equation}
\item The delay arising from the discretization of the PI controller
  and the Low-Pass Filter (LPF), with each introducing a time delay
  $\tau$. This delay is equivalent to the sampling time used in the
  discretization of the PI controller and LPF CTM transfer functions.
\end{enumerate}

Next, we consider the transfer function of each component in the control loop. First the PI controller transfer function is given by

\begin{equation}
  \label{eq:pi_controller_tf}
  H_p(s) = \frac{V_{p}(s)}{\Omega_e(s)} = k_p \frac{\tau_{iw} s +
    1}{\tau_{iw} s} e^{-s\tau}, 
\end{equation}
\begin{equation}
  \label{eq:tau_iw}
  \tau_{iw} =\frac{k_p}{k_i}
\end{equation}
where $k_p$ is the proportional gain, and $k_i$ is the integral gain.

The Pulse-Width-Modulator (PWM) component has the transfer function, $\SubText{H}{PWM}(s)$, which is given by,
\begin{equation}
  \label{eq:bldcm_chopper_tf}
  \SubText{H}{PWM}(s)= \frac{V_a(s)}{V_p(s)} = \frac{\SubText{V}{dc}}{\SubText{\tau}{pwm} s + 1}, 
\end{equation}
where $\SubText{\tau}{pwm}=0.5\SubText{f}{pwm}$, and
$\SubText{f}{pwm}$ is the modulation frequency. Typically,
$\SubText{f}{pwm}$ is much faster than the sampling frequency ($f_s=1 / \tau$). This allows $\SubText{H}{PWM}(s)$ to be reasonably approximated using $\SubText{H}{PWM}(s)\approx \SubText{V}{dc}$.

The motor electrical transfer function between the phase current and the terminal voltage is given by
\begin{equation}
  \label{eq:He-elec}
  \SubText{H}{elec}(s) = \frac{I_a(s)}{V(s)} = \frac{1/R}{\tau_e s + 1}, ~
\end{equation}
where $\tau_e=\frac{L}{R}$, while its mechanical transfer function, that is, the one between rotational speed and driving torque is given by

\begin{equation}
  \label{eq:H-mech}
  \SubText{H}{mech}(s) = \frac{\Omega_r(s)}{T_e(s)-T_l(s)} = \frac{1/B_m}{\tau_m s + 1},  
\end{equation}
where $\tau_m=\frac{J}{B_m}$.

The transfer function between the terminal voltage $V(s) = V_a(s) - V_b(s)$ and the shaft
rotational speed $\Omega_r(s)$, denoted by $H_m(s)$, can be obtained from \eqref{eq:bldc_state_space_model_2} using the system transfer function $\mat{H}(s) = \mat{C}_{2 \times 2}\left(s\mat{I} - \mat{A}_{2 \times 2} \right)^{-1} \mat{B}_{2 \times 2}$ and setting $T_l = 0$ (assuming that $B_m \approx 0$), as shown next,
\begin{equation}
  \label{eq:order2_bldcm_model_2}
  H_m(s) = \frac{\Omega_r(s)}{V(s)} = \frac{1/k_e}{\frac{R J}{k_e^2} \frac{L}{R} s^2 + \frac{R J}{k_e^2} s + 1}
\end{equation}
Defining the electrical and mechanical time constants as $\SubText{\tau}{elec}
= L / R$ and $\SubText{\tau}{mech} = R J / k_e^2$, respectively, we can
rewrite \eqref{eq:order2_bldcm_model_2} as
\begin{equation}
  \label{eq:order2_bldcm_model_3}
  H_m(s) =  \frac{1/k_e}{\SubText{\tau}{mech} \SubText{\tau}{elec} s^2 + \SubText{\tau}{mech} s + 1}
\end{equation}

Finally, the LPF has the transfer function 
\begin{equation}
  \label{eq:H_LPF}
  \SubText{H}{LPF}(s) = \frac{\Omega_f(s)}{\Omega_s(s)} = \frac{k_f}{\tau_fs+1} e^{-s\tau}
\end{equation}

From control systems theory we can easily see that the set-point tracking transfer function
$\SubText{H}{SP}(s)$ assuming the load torque input $T_l = 0$ is given by,
\begin{eqnarray}
  \label{eq:set_point_change_tf}
  \SubText{H}{SP}(s) &=& \frac{\Omega_r(s)}{\SubText{\Omega}{ref}(s)}\IEEEnonumber\\ 
                     &=& H_p(s) \SubText{H}{PWM}(s) H_m(s) H_{f}(s) e^{-s\tau} \left(\vphantom{e^{-s(2\tau + \tau_h)}}1 + H_p(s) \right. \IEEEnonumber\\ 
& \hphantom{=} &\hphantom{+}\times\left.\SubText{H}{PWM}(s) H_m \SubText{H}{LPF}(s) e^{-s(2\tau + \tau_h)}\right)^{-1}
\end{eqnarray}
where $H_{f}(s)$ is the transfer function  of the input speed profiler.

Similarly, taking the rotational reference speed $\SubText{\omega}{ref} = 0$ enables deriving the load disturbance transfer function, $\SubText{H}{load}(s)$, as follows,
\begin{eqnarray}
  \label{eq:load_disturbance_tf}
  \SubText{H}{load}(s) &=& \frac{\Omega_r(s)}{T_l(s)} \IEEEnonumber\\
                       &=& {\SubText{H}{mech}(s)}  \left(\vphantom{e^{-s(2\tau + \tau_h)}} 1+k_e \SubText{H}{elec}(s) \SubText{H}{mech}(s)  \right. \IEEEnonumber\\ 
                               & \hphantom{=} &+ H_p(s) \SubText{H}{PWM}(s)    \SubText{H}{LPF}(s) \SubText{H}{elec}(s) \IEEEnonumber\\
                                & \hphantom{=456} & \hphantom{+}\times\left.\SubText{H}{mech}(s) e^{-s(2\tau + \tau_h)}  \right)^{-1}
\end{eqnarray}

The continuous-time delay of the Hall sensor
$\tau_h$ can only be seen between two sampling times
$\tau$. Therefore, its contribution to the loop delay is given by
$\sup(\frac{\tau_h}{\tau})$. Looking at equations 
\eqref{eq:set_point_change_tf} and \eqref{eq:load_disturbance_tf}, we
can notice that, in both transfer functions, all delay contributions are
lumped together in the total time-delay 
\begin{equation}
  \label{eq:3}
  \SubText{\tau}{total} = m \tau
\end{equation}
with
\begin{equation}
  \label{eq:5}
  m=2+\sup(\frac{\tau_h}{\tau}).
\end{equation}
\subsection{Summary and Discussion}
\label{sec:summary-discussions}
The preceding developments in this section aimed at taking into
account all the sources of delays induced by the various modules in
the control loop of the BLDCM and arriving to an LTI-RTDS of the form
\eqref{eq:1}. This was accomplished by deriving the set-point tracking
and load-disturbance rejection transfer functions, respectively, in
\eqref{eq:set_point_change_tf} and
\eqref{eq:load_disturbance_tf}. Converting those transfer functions
into the state-space LTI-RTDS format of \eqref{eq:1} is
straightforward but not needed at this point, since the current goal
is to use the method outlined in Section \ref{sec:methodology}, which
proceeds starting from the characteristic polynomial in
\eqref{eq:det-ch-poly-1}. The next step is therefore to extract the
characteristic polynomial corresponding to the transfer functions in
\eqref{eq:set_point_change_tf} or
\eqref{eq:load_disturbance_tf}. It is to be noted that $\SubText{\tau}{total}$ in those
transfer functions represents the actual delay of the system, which
was labelled simply as $\tau$ in Section~\ref{sec:methodology}.

The next section will pursue the quest of computing $\tau_{\mathrm{max}}$ for a system whose transfer functions are given by \eqref{eq:set_point_change_tf} and \eqref{eq:load_disturbance_tf}.
%############################################################################
\section{Effect of Delay on the Stability}
\label{sec:effect-delay-stabl}
This section aims at applying the method outlined in Section
\ref{sec:methodology} to compute $\SubText{\tau}{max}$ of the
delay-based model of the BLDCM developed in Section \ref{sec:delay-based-modell}.

As a first step towards towards this goal, we first note that  the denominators of the above derived
transfer functions \eqref{eq:set_point_change_tf} and
(\ref{eq:load_disturbance_tf}) are quasi-polynomials in $s$, which
take the same form as the characteristic polynomial of the LTI-RTDS in
\eqref{eq:det-ch-poly-1}. Therefore, the steps described in Section
\ref{sec:methodology} to determine the maximum delay of stable
operation of the LTI-RTDS can be employed, with slight modifications, to
determine $\tau_{\max}$: the maximum allowable delay for $\SubText{\tau}{total}$ under
stable operation. Denoting the denominator
quasi-polynomials in \eqref{eq:set_point_change_tf} and
\eqref{eq:load_disturbance_tf} by
$\textnormal{CE}_{\textnormal{SP}}(s,\SubText{\tau}{total})$, and
$\textnormal{CE}_{\textnormal{load}} (s,\SubText{\tau}{total})$ while using the delay $\SubText{\tau}{total}$ defined in \eqref{eq:3} and \eqref{eq:5}, we get
\begin{multline}
  \label{eq:set_point_change_ce}
  \textnormal{CE}_{\textnormal{SP}} (s,\SubText{\tau}{total})(s)= k_s (\tau_{iw}s+1) e^{-s \SubText{\tau}{total}} \\+ \tau_{iw}s (\tau_l
  s+1) (\tau_{f1}s+1)
\end{multline}
\begin{multline}
  \label{eq:load_disturbance_ce}
  \textnormal{CE}_{\textnormal{load}} (s,\SubText{\tau}{total}) (s) = k_l (s\tau_{iw}+1) e^{-s \SubText{\tau}{total}} + s k_m (s\tau_{f1}+1)  \\
           +  s k_n (s\tau_e+1) (s\tau_m+1) (s\tau_{f1}+1) 
\end{multline}
where $k_s = \frac{k_p \SubText{V}{dc}}{k_e}$, $k_l = \frac{k_p k_e k_{f1}
\SubText{V}{dc}}{R}$, $k_m = \tau_{iw} \frac{k_e^2}{R}$ and $k_n = B_m \tau_{iw}$.

Next, we proceed with Step 1 in the method described in Section
\ref{sec:methodology} to find $\tau_{\max}$. We will limit the
following analysis to $\textnormal{CE}_{\textnormal{load}}
(s,\SubText{\tau}{total}) (s) $ noting that
$\textnormal{CE}_{\textnormal{SP}}(s,\SubText{\tau}{total})$ can  be
treated in a like manner.

Using the Rekasius mapping \eqref{eq:mapping-1} in
\eqref{eq:load_disturbance_ce} results in transforming \eqref{eq:load_disturbance_ce} into, 

\begin{equation}
  \label{eq:load_disturbance_test_2}
  \widebar{\textnormal{CE}}_{\textnormal{load}}(s, T) = \sum_{i=0}^5q_i(T) s^i 
\end{equation}
where,
\begin{eqnarray*}
  \label{eq:load_disturbance_test_3}
  q_5(T) &=& k_n \tau_e \tau_m \tau_{f1} T   \\
  q_4(T) &=& k_n \tau_e \tau_m \tau_{f1} + (k_n \tau_e \tau_m + k_n \tau_e
        \tau_{f1} + k_n \tau_m \tau_{f1}) T   \\ 
  q_3(T) &=& k_m \tau_e \tau_m + k_n \tau_e \tau_{f1} + k_n \tau_m \tau_{f1}\\
            & \hphantom{=}& + (k_m \tau_{f1} + k_n \tau_e + k_n \tau_m + k_n \tau_{f1}) T   \\
  q_2(T) &=& k_m \tau_{f1} + k_n \tau_e + k_n \tau_m + k_n \tau_{f1} \\
            & \hphantom{=}& +  (k_m + k_n - k_l \tau_{iw}) T   \\
  q_1(T) &=& k_l \tau_{iw} + k_m + k_n - k_e T, \\
  q_0(T) &=& k_l  
\end{eqnarray*}

The rest of steps in Section \ref{sec:methodology}, starting with step
2, can be automated and implemented in a Matlab script culminating
with $\tau_{\max}$.

At this point, few remarks are worthy of note in order to highlight
the significance of $\tau_{\max}$ computed by the above procedure.
\begin{itemize}
\item Given that the delay-free speed controller (\SubText{\tau}{total} = 0) is nominally stable by design, it follows that $\tau_{\max}$ computed using the above steps represents an upper bound on $\SubText{\tau}{total}$ beyond which the system becomes unstable. In other words, $\tau_{\max}$ is the maximum delay that the closed-loop system can tolerate before the roots of its characteristic quasi-polynomial cross to the RHP (rendering the system unstable) as $\SubText{\tau}{total}$ is increased above $0$. 
\item The value for $\tau_{\max}$ calculated by the above steps depends solely on the BLDC motor and speed controller parameters, and is independent of the motor operating conditions, e.g. the rotational speed of the motor. This fact makes the condition $\SubText{\tau}{total} < \tau_{\max}$ a necessary condition for the stability that is given \textit{a priori} independent of the operating conditions.
\item The value of \SubText{\tau}{total} (which depends on the actual operating conditions of the motor) and its proximity to $\tau_{\max}$ can be used to serve as a measure for the stability of the speed controller. For example, the further $\SubText{\tau}{total}$ is from $\tau_{\max}$ the closer the system is to its stable delay-free condition.
\item Given a nominal set of operating conditions, a controller design with bigger $\tau_{\max}$ is more robust to changes in the closed-loop delay.
\end{itemize}

%############################################################################
\section{Effect of Controller Parameters on the Stability}
\label{sec:effect-controller-stabl}

The goal in this section is employ the procedure developed in Section
\ref{sec:effect-delay-stabl} for computing $\SubText{\tau}{max}$ as a
lens, so to speak, that enables viewing the various commonly used PI
tuning rules from a totally different angle. More specifically, we examine the impact of the choice of the controller parameters on $\tau_{\max}$. This task will be carried out in several steps. 
\begin{itemize}
\item In the first step, we will consider some of the widely used PI tunning rules, and examine their choices for the PI controller parameters ($k_p$ and $k_i$) through the lens of their impact on $\SubText{\tau}{max}$. This step is given in Section \ref{sec:effect_of_tuning_rules}.
\item Next, we will let those parameters vary continuously, within
  reasonable ranges, and plot, in Section \ref{sec:effect_of_kp_ki},
  the corresponding values for $\SubText{\tau}{max}$, where we find
  new insights that, to the best of the authors' knowledge, remained hitherto unknown.
\item Finally, in Section \ref{sec:effect_of_lpf}, we study the effect
  of the LPF cutoff frequency, $\omega_f$, on $\tau_{\max}$.
\end{itemize}

Before proceeding further, we need to present the operating conditions and the basic setup established for this study. 

The motor chosen to conduct this study is the Beijing BL3056 which comes with the TI Stellaris RDK-BLDC design kit (which we refer to as the TI controller). The main motor parameters are given in Table \ref{table:bl3056_parameters}, and the TI controller parameters, for the LPF and PI controller, are given in tables \ref{table:pi_lpf_parameters} and \ref{table:controller_comparison_table} respectively. The reader is referred to \cite{TiBldcRdk10:_TI_BLDC_RDK_users_guide} for additional details. We also considered the load disturbance rejection response, through its quasi-polynomial, to calculate $\tau_{\max}$. The control input is the disturbance torque in N.m and the output is the BLDCM shaft angular speed in $rad/s$. Finally, the total loop delay $\SubText{\tau}{total}$ was set to $3.7\textnormal{ms}$ (equivalent to a rotational speed of $6000 ~\textnormal{RPM}$ and $\tau = 1\textnormal{ms}$).

\begin{table}[htbp]
  \centering
  \caption{Beijing BLDCM BL3056 Parameters.} 
  \begin{tabular}[c] { c c c c c c c}
    \toprule
          $R$    &   $L$  &   $J$      &  $k_t$  &  $k_e$  & $\SubText{\tau}{elec}$ & $\SubText{\tau}{mech}$ \\
          $\Omega$    &  mH  &   $\textnormal{g.cm}^2$      &  N.A/m  &  V/RPM  &  ms & ms \\
          \midrule\midrule
         2.3    & 0.56 & 16.0     & 0.0223  & 0.00234 &     0.24      &    7.4      \\  \bottomrule
  \end{tabular}
  \label{table:bl3056_parameters}
\end{table}

\begin{table}[htbp]
  \centering
  \caption{Parameters of TI Stellaris RDK-BLDC Speed Controller}
\begin{tabular}[c]{c c c }
    \toprule
        $k_{f}$  & $\tau_{f}$ (ms) & $\omega_{f}$ (rad/s)\\\midrule
          1.0    &    3.48     &     287.7     \\\bottomrule
  \end{tabular}
  \label{table:pi_lpf_parameters}
\end{table}

\subsection{Effect of PI Controller Tuning Rules on $\SubText{\tau}{max}$}
\label{sec:effect_of_tuning_rules}
We analyze in this section the relationship between $\SubText{\tau}{max}$ and
various popular PI controller tuning rules including Ziegler-Nichols
(Z-N), Chien-Hrones-Reswick (CHR) and the the methods based on the
integral error criteria (Integral of Square Error - ISE, Integral of
Absolute Value Error - IAE, Integral Time Squared Error - ISTE and
Integral Time Absolute Error - ITAE) \cite{Xue07:_PID_controller_design}. 

Table \ref{table:controller_comparison_table} shows the values of $k_p$ and $k_i$, and the corresponding values for $\tau_{\max}$, obtained for each one of those tunning rules, with the values of the LPF set to those given in Table \ref{table:pi_lpf_parameters}. Note that several tuning rules of those listed in Table \ref{table:controller_comparison_table} have two different values for both $k_p$ and $k_i$. Those two sets of values correspond to whether the tunning rule is being optimized for load disturbance rejection or set point tracking responses, marked by \textdagger or \textdaggerdbl, respectively. The table also shows, in the first row, the values for $k_p$ and $k_i$ selected by the TI controller, as well as the response obtained by the custom rule of the TI controller.

\begin{table}[htbp]
  \begin{threeparttable}[b]
  \centering
  \caption{Effect of Various Tuning Rule on $\SubText{\tau}{max}$}
  \begin{tabular}[c]{ c c c c c }
    \toprule
    \multirow{3}{*}{\parbox{10mm}{\centering Tuning\\ Rule}} &  \multicolumn{4}{c}{Parameters values}\\\cmidrule{2-5}
                   & $k_p\times 10^{-3}$      & $k_i\times 10^{-6}$      & $\tau_{iw}=\frac{k_p }{ k_i}$ (ms) & $\SubText{\tau}{max}$ (ms)   \\\toprule
                   TI \textasteriskcentered & 0.122 & 0.366 & 333   & 4274 \\\midrule\midrule
                   CHR \textdagger          & 1.024 & 65.43 & 15.65 & 12.8 \\\midrule
                   ISE  \textdagger         & 0.669 & 20.10 & 33.28 & 18.9 \\\midrule
                   ISTE \textdagger         & 0.527 & 26.87 & 19.61 & 16.7 \\\midrule\midrule
                   Z-N \textdaggerdbl       & 1.536 & 117.9 & 13.03 & 5.2  \\\midrule
                   CHR \textdaggerdbl       & 1.024 & 142.5 & 7.188 & 8.3  \\\midrule
                   ISE \textdaggerdbl       & 1.566 & 132.0 & 11.87 & 4.9  \\\midrule
                   ISTE \textdaggerdbl      & 1.158 & 134.3 & 8.624 & 7.5  \\\midrule
                   IAE \textdaggerdbl       & 1.160 & 135.9 & 8.532 & 7.4  \\\midrule
                   ITAE \textdaggerdbl      & 1.472 & 135.9 & 10.83 & 5.3  \\\bottomrule
  \end{tabular}
  \begin{tablenotes}
  \item [\textasteriskcentered] The tuning parameters set by the TI Stellaris kit.
  \item [\textdagger] Tunning rules for load disturbance rejection response.
  \item [\textdaggerdbl] Tunning rules for set-point tracking response.
  \end{tablenotes}
  \label{table:controller_comparison_table}
   \end{threeparttable}
\end{table}

\begin{figure}[hbtp]
  \centering
  \includegraphics[width=0.4\textwidth]{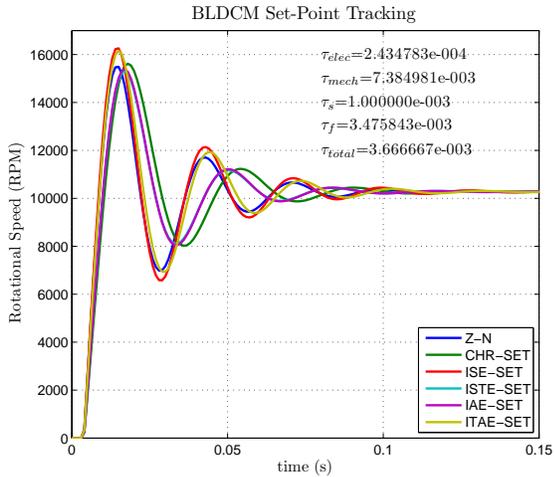}
  \caption{PI controller Set-Point Tracking Response for Various Tuning Rules.}
  \label{fig:pi_tuning_rules_set}
\end{figure}

\begin{figure}[hbtp]
  \centering
  \includegraphics[width=0.4\textwidth]{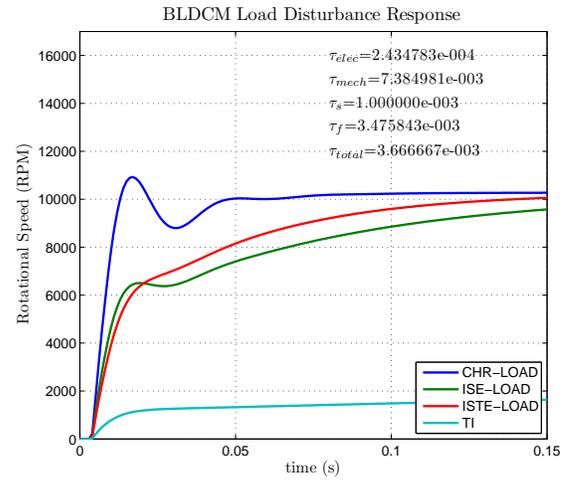}
  \caption{PI controller Load Disturbance Response for Various Tuning Rules.}
  \label{fig:pi_tuning_rules_load}
\end{figure}

To visualize the performance obtained from each of the above tunning rules, Figs. \ref{fig:pi_tuning_rules_set} and \ref{fig:pi_tuning_rules_load} present the transient response of the system. Fig. \ref{fig:pi_tuning_rules_set} groups the result corresponding to tunning rules optimized for set point tracking responses, whereas Fig. \ref{fig:pi_tuning_rules_load} shows the results obtained for the rules optimized for load disturbance rejection. 

The foremost remarks seen from the above results can be summarized by the following points.
\begin{itemize}
\item The tunning rules optimized for load disturbance rejection produce behaviours more robust to delays compared to those optimized for set point tracking response. It should be observed too that the former set results in larger values for $\tau_{\max}$ than the latter set.
\item The TI commercial controller tuning values of Table
  \ref{table:pi_lpf_parameters} result in an excessively large $\SubText{\tau}{max}$ compared to the other methods. Hence, the TI controller should be very robust regarding the loop delay $\SubText{\tau}{total}$. However, this highly desirable property comes at the expense of a very sluggish transient response as shown in Fig. \ref{fig:pi_tuning_rules_load}.
\end{itemize}

The above remarks are indeed in line with the rationale behind the proposed method, which is premised on the notion that design methodologies that yield larger values for $\SubText{\tau}{max}$ produce more robust behavior that is less sensitive to loop delays.


\subsection{Effect of Widely Varying $k_p$ and $k_i$ on $\SubText{\tau}{max}$}
\label{sec:effect_of_kp_ki}
In this section we further analyze the impact of the speed controller design choices on the value of $\tau_{\max}$, through sweeping the
values of $k_p$ and $k_i$, for some selected values of $\omega_f$, and calculating $\tau_{\max}$ using the method proposed in section \ref{sec:effect-delay-stabl}. 

\subsubsection{High $\omega_f$}
\label{sec:high-omega_f}
\begin{figure}[hbtp!]
  \centering
  \includegraphics[width=0.4\textwidth]{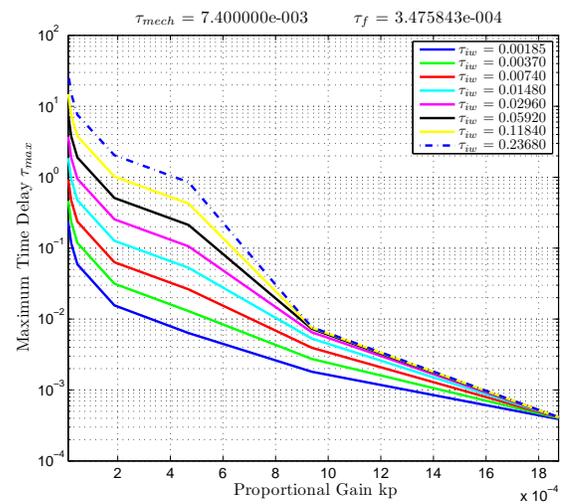}
  \caption{Effect of PI controller on $\tau_{max}$ ($\omega_f = 2877 ~rad/s$).}
  \label{fig:pi_kp_effect_delay_noLPF}
\end{figure}

In the first set of experiments, we minimize the effect of the LPF by setting $\omega_f$ to be one order of magnitude bigger than the value in table \ref{table:pi_lpf_parameters} ($\omega_f = 2877$rad/s). Fig. \ref{fig:pi_kp_effect_delay_noLPF} shows $\tau_{\max}$ corresponding to sweeping $k_p$, in a wide range of values, while setting $\tau_{iw}$ (by adjusting $k_i$) to values ranging from $0.25\SubText{\tau}{mech}$ to $32\SubText{\tau}{mech}$.

It is important to note here that at high $k_p$ values (about
$1.8\times 10^{-3}$), the smaller values for $\SubText{\tau}{max}$
suggest that the controller becomes more sensitive to delays in the
control loop. We can also see that changes in $\tau_{iw}$ do not
significantly affect the maximum loop time-delay
$\SubText{\tau}{max}$, in this range. Consequently, around this range, $k_p$
dominates the controller behavior and robustness to delays.

On the other hand, at lower values of $k_p$, $\SubText{\tau}{max}$ can
be many orders of magnitude bigger than its value at high
$k_p$. Consequently, the controller is much more robust to delays at
low $k_p$. It is important to notice that at this range, for the same
$k_p$, $\SubText{\tau}{max}$ can vary by more than two orders of
magnitude depending on the value of $\tau_{iw}$. Therefore, at this
range, the controller designer has two degrees of freedom ($k_p$ and
$\tau_{iw}$) to design a controller with a proper trade-off between
speed and robustness to delays.

\subsubsection{Medium and small $\omega_f$}
\label{sec:effect_of_lpf}
\begin{figure}[hbtp]
  \centering
  \includegraphics[width=0.4\textwidth]{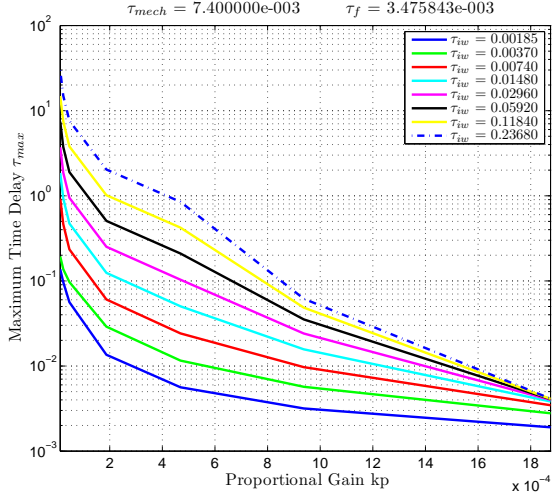}
  \caption{Effect of PI controller on $\tau_{max}$ ($\omega_f = 287.7 ~rad/s$).}
  \label{fig:pi_kp_effect_delay_withLPF_1}
\end{figure}

Here we analyze the effect of the LPF cutoff
frequency $\omega_{f}$ on the robustness of the speed controller to
time-delays. First, we set $\omega_{f}$ to its TI controller value
of table \ref{table:pi_lpf_parameters} and repeat the calculations of
section \ref{sec:effect_of_kp_ki}. As we can see in Fig.
\ref{fig:pi_kp_effect_delay_withLPF_1}, at this cutoff frequency and at high
$k_p$, changes in $\tau_{iw}$ affect $\SubText{\tau}{max}$ more than
for the previous case when $\omega_{f}$ was large. On the other hand,
at low $k_p$, the smaller $\omega_{f}$ causes almost no change on 
$\SubText{\tau}{max}$. 

\begin{figure}[hbtp]
  \centering
  \includegraphics[width=0.4\textwidth]{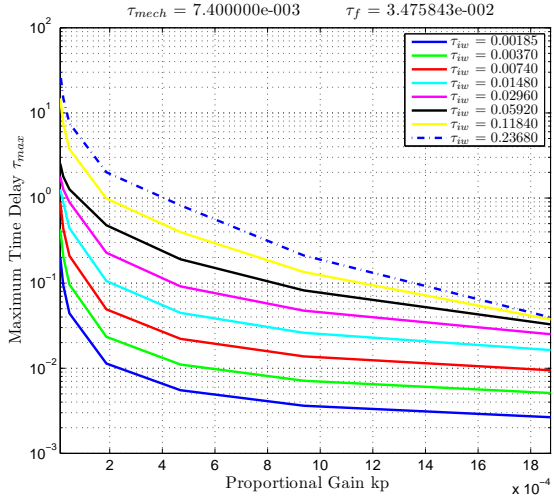}
  \caption{Effect of PI controller on $\tau_{max}$ ($\omega_f = 28.77 ~rad/s$).}
  \label{fig:pi_kp_effect_delay_withLPF_2}
\end{figure}

Decreasing $\omega_{f}$ further to $10 \%$ the value in table
\ref{table:pi_lpf_parameters} causes $\SubText{\tau}{max}$ to become even more
sensitive to $\tau_{iw}$ at high $k_p$, as shown in figure
\ref{fig:pi_kp_effect_delay_withLPF_2}. Actually, $\SubText{\tau}{max}$
becomes significantly bigger indicating that the controller becomes less
sensitive to delays in the control loop with smaller LPF cutoff
frequencies. However, $\SubText{\tau}{max}$ values at low $k_p$ remain
almost unchanged.

%############################################

\section{Experimental Results}
\label{sec:experimental_results}
This section demonstrates experimentally the advantage of the proposed stability analysis under different operating conditions. The approach followed in the presentation of this section is to juxtapose the stability criteria established by the proposed method, namely, the requirement that $\SubText{\tau}{total} < \tau_{\max}$ for stable operation, and the classical stability conditions based on the phase and gain margins obtained through the Nyquist plot. This process is carried out for speed controllers designed by a subset of the tunning tuning rules given in Table \ref{table:controller_comparison_table} for different operating conditions of the rotational speed $\omega_r$.

It should be noted, however, that the criteria based on the classical methods need to be repeated for all desired operating conditions, whereas in the proposed method $\tau_{\max}$ is independent of the operating conditions and therefore is more convenient to use.

This section presents both simulation and experimental results whose setup is described briefly in the next subsection.

\subsection{Simulation and Experimental Setup}
\label{sec:simul-exper-setup}
The transient time-domain simulation results were obtained by constructing a Simulink model for the BLDC motor and the PI controller and LPF.  Those transient simulations show the simulation results for the continuous-time model (CTM) with and without delay and for the discrete-time model (DTM) with delay.  The experimental results were obtained using
a Stellaris BL3056 BLDC motor control reference design kit (RDK-BLDC) and a Xilinx University program Virtex II Pro Development kit. The generator stator was connected to various resistive loads through an array of electronically controlled power switches. This setup allowed us to quickly connect and disconnect loads to the generator creating accurate transient torque scenarios with good accuracy and repeatability.

\subsection{Description of the Experiments}
\label{sec:experimental_results_nyquist}
The rows in Table \ref{table:tds_comparison_table} summairze the results obtained for the TI speed controller, as well as three of the tunning rules given earlier in Table \ref{table:controller_comparison_table} under two operating conditions of the rotational speeds $\omega_r= 6000 $ and $1000$ RPM, which are about $75 \%$ and $10 \%$ of the BL3056 rated speed, respectively. Those values of $\omega_r$ correspond to values of $\tau_h$ of 1.67 and 10 ms, respectively. Table \ref{table:tds_comparison_table} also provides the Phase Margin (PM) obtained in each tuning rule and under the two operating conditions. The table illustrates the agreement between the marginal stability criteria in the classical method (PM $> 0^{\circ}$) and the stability criteria in the proposed method ($\SubText{\tau}{total} < \tau_{\max}$). However, the classical method based on the PM does not provide the amount of delay in $\SubText{\tau}{total}$ before the system becomes unstable.

\begin{table*}[htbp!]
  \centering
  \caption{Stability Conditions under Two Different Operating Conditions }
  \begin{threeparttable}
    \begin{tabular}[c]{c c   cc c  c   c c }
      \toprule
      \multirow{3}{*}{\ParComp{20mm}{Tuning Method}}  &\multirow{3}{*}{$\tau_{\max}$} & \multicolumn{5}{c}{Operating Condition $\omega_r$, \SubText{\tau}{total}}                                   \\\cmidrule{3-7} 
                                                         &                                              & \multicolumn{2}{c}{6000 RPM, \SubText{\tau}{total} =3.7 ms}                                        &  & \multicolumn{2}{c}{1000 RPM, \SubText{\tau}{total} = 12.0 ms} \\\cmidrule{3-4} \cmidrule{6-7}
                                                         &                                              & $\SubText{\tau}{total} < \tau_{\max}$           & PM            & &   $\SubText{\tau}{total} < \tau_{\max}$ &PM \\\midrule\midrule
              TI                                        &              4274                                                     & TRUE                      & 82.7         &                          & TRUE      & 82.1\\\midrule
              CHR\textdagger                 &              12.8                                                       & TRUE                     & 42.4          &                          & TRUE      & 9.87\\\midrule
              ISE\textdagger                   &              18.9                                                       & TRUE                     & 59.9          &                          & TRUE      & 46.6\\\midrule
              Z-N                                    &                5.2                                                       & TRUE                     & 24.8          &                          & \textbf{FALSE}     & \textbf{-35.2}\\
      \bottomrule
    \end{tabular}
    \begin{tablenotes}
    \item [\textdagger] Tunning rules for optimized load distrubance rejection response.
  \end{tablenotes}
  \end{threeparttable}
  \label{table:tds_comparison_table}
\end{table*}

\subsubsection{Stability Charactreistics of the TI Conroller}
\label{sec:stab-char-ti}
Figs. \ref{fig:nyquist_TI_1k}-~\ref{fig:nyquist_TI_3} show the Nyquist plots and the transient responses for the TI controller under three rotational speeds $\omega_r = 1000$, $5$ and $3$ RPM, respectively. 

For $\omega_r = 1000$ RPM, $\tau_h \approx 1$ms. Therefore, $\SubText{\tau}{total} << \tau_{\max}$, implying that the TI controller should be very stable with no oscillations. This fact is confirmed by the transient response shown in Fig. \ref{fig:nyquist_TI_1k} which shows the three transient responses (CTM with and without delay, and DTM with delay) to conicide perfectly. However, this desired stability response comes at the expense of a very sluggish behaviour, as shown by the measured response in Figs. \ref{fig:exp_speed_trans_6000} and \ref{fig:exp_torque_trans_3000}, which show  TI controller transient response when the reference speed change from $1000 ~RPM$ to $6000 ~RPM$ (no load condition) and the load torque change from $0\textnormal{ ~mN} \cdot \textnormal{m}$ to $27.8 \textnormal{ ~mN} \cdot \textnormal{m}$, respectively.

On the other hand, for $\omega_r=5$ RPM we have a $\SubText{\tau}{total}=2$, which is slightly less than half of $\tau_{\max}$. As Fig. \ref{fig:nyquist_TI_5}, at this speed the system starts approaching the stability boundary as evidenced by the oscillations in the transient responses.

Finally, at $\omega_r = 3$ RPM,  $\SubText{\tau}{total}=3.34$, approaching the $\tau_{\max}$ limit of stable operation. As shown in Fig. \ref{fig:nyquist_TI_3} the response becomes increasingly oscillatory.

\subsubsection{Stability of Conservative and Agressive Tuning Rules}
\label{sec:experimental_results_}
\begin{figure}[hbtp]
  \centering
  \includegraphics[width=0.45\textwidth]{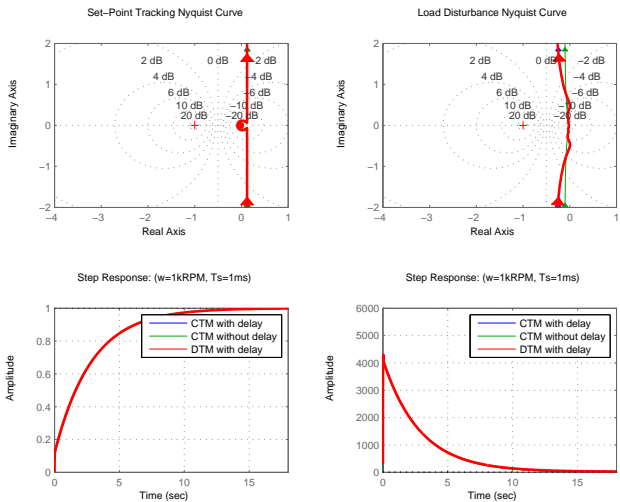}
  \caption{Delay Effect Using TI controller tuning values at
    $\omega_r=1 ~kRPM$.}
  \label{fig:nyquist_TI_1k}
\end{figure}

\begin{figure}[hbtp]
  \centering
  \includegraphics[width=0.45\textwidth]{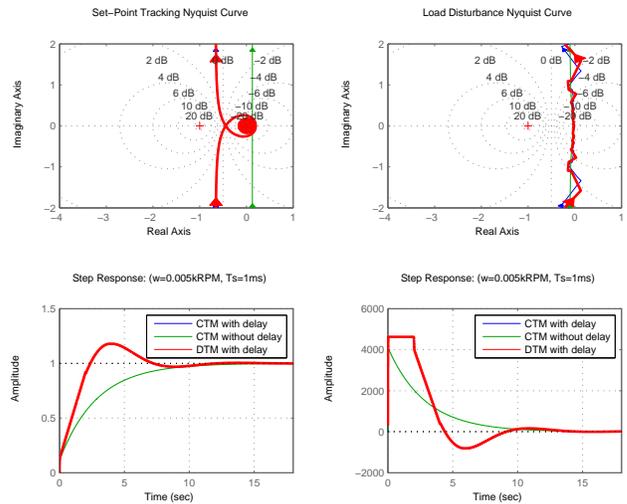}
  \caption{Delay Effect Using TI controller tuning values at $\omega_r=5 ~RPM$.}
  \label{fig:nyquist_TI_5}
\end{figure}

\begin{figure}[hbtp]
  \centering
  \includegraphics[width=0.45\textwidth]{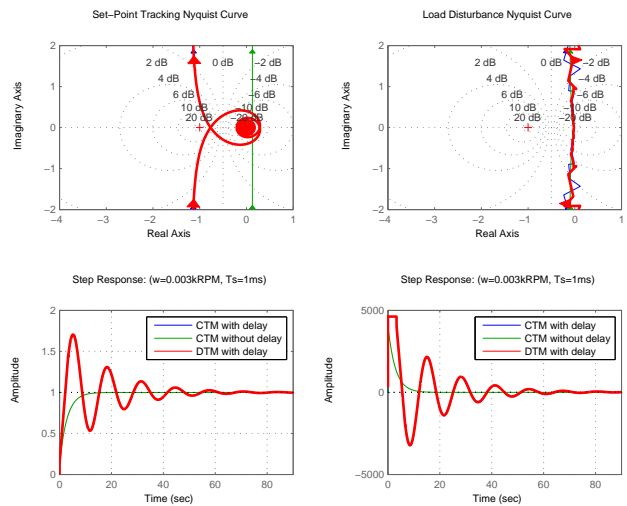}
  \caption{Delay Effect Using TI controller tuning values at $\omega_r=3 ~RPM$.}
  \label{fig:nyquist_TI_3}
\end{figure}

Figs. \ref{fig:nyquist_TI_1k}, \ref{fig:nyquist_TI_5} and
\ref{fig:nyquist_TI_3} show the Nyquist plot and the step response of
the TI controller at rotational speeds $\omega_r$ of $1000 ~RPM$, $5
~RPM$ and $3 ~RPM$ respectively. At $1000 ~RPM$ ($\tau_h = 10ms$), we
have $\SubText{\tau}{total} << \tau_{max}$ which implies the TI 
controller should be stable and show low sensitivity to
$\omega_r$. These results are confirmed by the similar gain margin for
$1000 ~RPM$ and $6000 ~RPM$ in table \ref{table:tds_comparison_table}
and no sign of oscillatory behavior in figure
\ref{fig:nyquist_TI_1k}. The small $\SubText{\tau}{total}$ and $\tau_h$, compared to
$\SubText{\tau}{max}$, cause the three graphics coincide in figure \ref{fig:nyquist_TI_1k}.

At $\omega_r = 5 ~RPM$ ($\SubText{\tau}{total} \approx 2ms$),
$\SubText{\tau}{total}$ is just a little less that half of $\SubText{\tau}{max}$. 
It is straightforward to see in figure \ref{fig:nyquist_TI_5} that the
system starts approaching the stability boundary as evidenced by the 
oscillation on its step response. Beyond $\SubText{\tau}{total} = 0.5 \cdot
\tau_{max}$ it quickly approaches instability becoming fully unstable
at $\SubText{\tau}{total} = \tau_{max}$. At $\omega_r = 3 ~RPM$ 
($\SubText{\tau}{total} = 3.34ms$), as $\SubText{\tau}{total}$ approaches $\SubText{\tau}{max}$, the step
response oscillatory behavior shown in figure \ref{fig:nyquist_TI_3}
has significantly increased.

\begin{figure}[hbtp]
  \centering
  \includegraphics[width=0.45\textwidth]{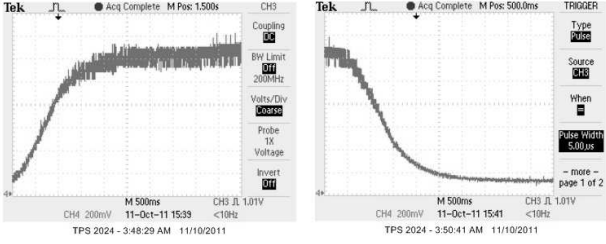}
  \caption{Speed transient from 1000 to 6000 ~RPM at no load.}
  \label{fig:exp_speed_trans_6000}
\end{figure}

\begin{figure}[hbtp]
  \centering
  \includegraphics[width=0.45\textwidth]{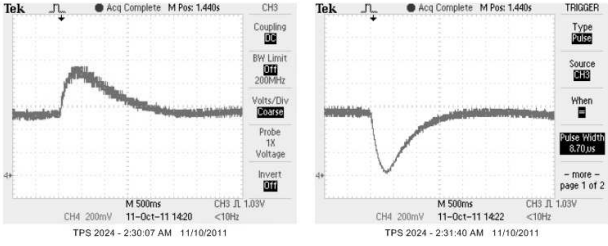}
  \caption{Torque transient test at 3000 ~RPM (0 - 27.8 m NM - 0).}
  \label{fig:exp_torque_trans_3000}
\end{figure}

Figures \ref{fig:exp_speed_trans_6000} and
\ref{fig:exp_torque_trans_3000} show the It confirms the sluggish response of
the TI controller to set-point tracking and load disturbance rejection
responses.

In summary, the TI controller $k_p$ and $k_i$ values produce a controller
capable of handling total loop time-delays of up to 
$2.0 s$. However, its robustness to loop time-delay
comes at the expense of very slow dynamical responses. It is highly
desirable to tune PI controllers to robustness and stability but still
be able to achieve fast dynamical response.
$\newline$

\subsubsection{Stability of Fast And Stable Tuning Rules}
\label{sec:experimental_results_fast}
The objective of this section is to study the stability conditions of the remaining three rules tuned to fast response and how $\SubText{\tau}{max}$ correlates to them. Here we compare two tuning rules: CHR-load and ISE-load. The popular Ziegler-Nichols (Z-N) method is well known for producing oscillatory responses. Looking at the gain and phase margins of table \ref{table:tds_comparison_table}, we can see the Z-N tuning values result in a $\SubText{\tau}{total}$ that is too close to $\SubText{\tau}{max}$ and will not be considered here.

\begin{figure}[hbtp]
  \centering
  \includegraphics[width=0.45\textwidth]{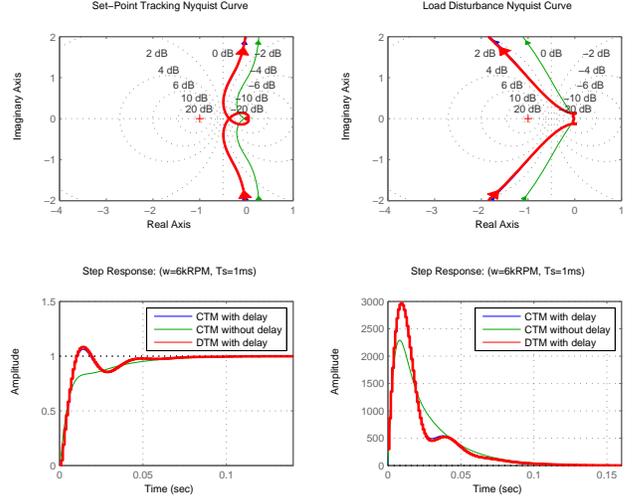}
  \caption{Delay Effect Using CHR-load tuning values at $\omega_r=6 ~kRPM$.}
  \label{fig:nyquist_chrload_6k}
\end{figure}

\begin{figure}[hbtp]
  \centering
  \includegraphics[width=0.45\textwidth]{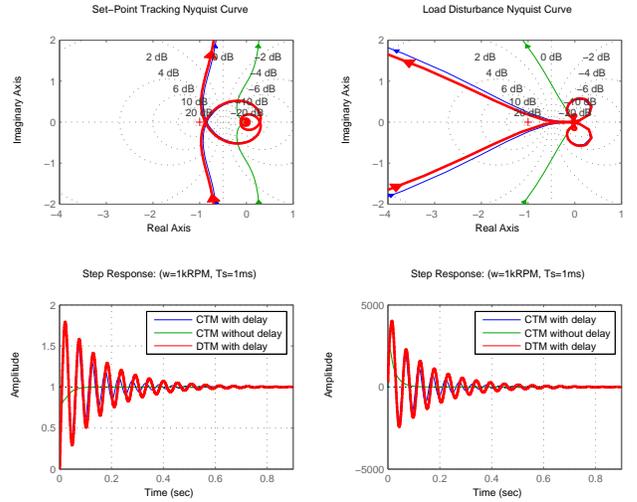}
  \caption{Delay Effect Using CHR-load tuning values at $\omega_r=1 ~kRPM$.}
  \label{fig:nyquist_chrload_1k}
\end{figure}

Among the classical methods tested, the CHR, for optimized load disturbance rejection,  of Figs. \ref{fig:nyquist_chrload_6k} and \ref{fig:nyquist_chrload_1k} provided the best compromise between smooth and fast response at $\omega_r = 6000 ~RPM$. At this speed, $\SubText{\tau}{total}$ is only $30 \%$ of ${\tau}_{\max}$ ($\tau_{\max} = 12.8 ms$) with a rise time of about $10 ms$
and overshoot of $30 \%$. For te same BLDCM and LPF, the CHR tuning, also for optimized load disturbance rejection, is about $700$ faster than the TI controller.
As we increase $\tau_h$ (by decreasing $\omega_r$ to $1000 ~RPM$), we
have $\SubText{\tau}{total} \approx \tau_{\max}$ causing the system to behave with a
significant oscillatory step response at $\omega_r = 1000 ~RPM$.

\begin{figure}[hbtp]
  \centering
  \includegraphics[width=0.45\textwidth]{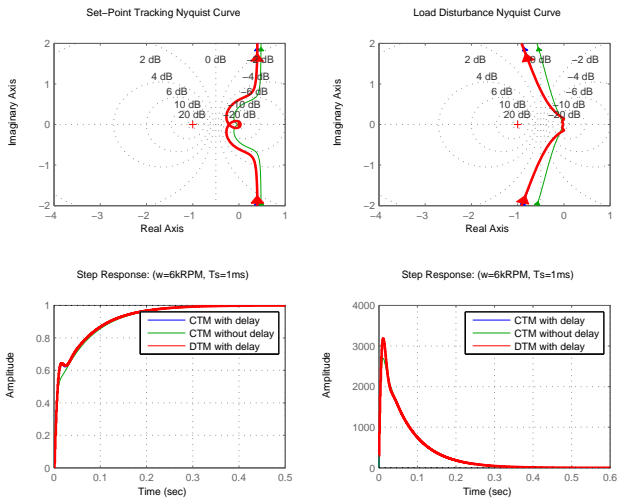}
  \caption{Delay Effect Using ISE-load tuning values at $\omega_r=6 ~kRPM$.}
  \label{fig:nyquist_iseload_6k}
\end{figure}

\begin{figure}[hbtp]
  \centering
  \includegraphics[width=0.45\textwidth]{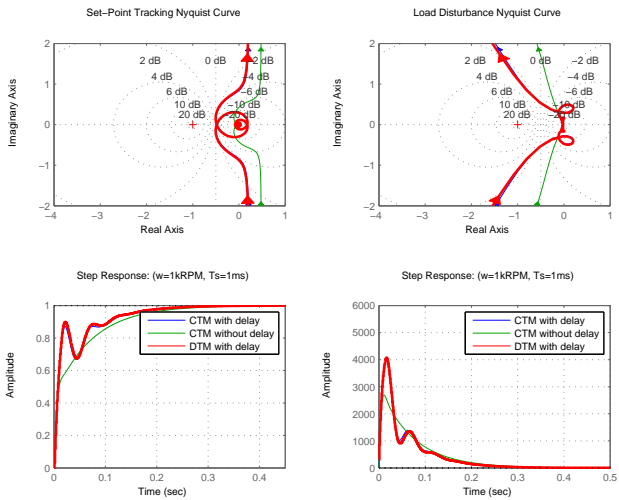}
  \caption{Delay Effect Using ISE-load tuning values at $\omega_r=1 ~kRPM$.}
  \label{fig:nyquist_iseload_1k}
\end{figure}

With a high $\SubText{\tau}{max}$ value of $18.9 ms$, the $ISE$ tuning method, optimized for load disturbance rejection, is
the most stable and robust to time-delays among all methods tested. At $\omega_r
= 6000 ~RPM$, its $\SubText{\tau}{total}$ is less than $20 \%$ of $\SubText{\tau}{max}$
resulting in a step response with no overshoot as shown in Fig.
\ref{fig:nyquist_iseload_6k}. This smoother response comes at the expense of
a rise time that is $12$ times longer compared to CHR (load) tuning
($120 ms$ for ISE (load) compared to $10 ms$ of CHR (load). With a $48
\%$ overshoot at $\omega_r = 1000 ~RPM$ ($\SubText{\tau}{total} \approx 0.6 \cdot
\tau_{max}$), the ISE-load step response in figure
\ref{fig:nyquist_iseload_1k} starts to show an oscillatory step response.

%############################################################################

%############################################################################
\section{Conclusions}
\label{sec:conclusions}
This work presented an effective algebraic technique to analyze the
asymptotic stability of a BLDCM speed controller with strong
time-delays in the control loop. We have shown that given a BLDCM, the
$k_p$ and $k_i$ PI controller parameters, and the LPF cutoff frequency
$\omega_{f1}$, there is a maximum time-delay $\SubText{\tau}{max}$ beyond which
the total total closed loop time-delay $\SubText{\tau}{total}$ causes the system to
become unstable and presented a method to calculate
$\SubText{\tau}{max}$. Experimental results confirm the method is more accurate
and less conservative than previously reported ones.  

Based on the mathematical definitions of $\SubText{\tau}{max}$ and $\SubText{\tau}{total}$, and
confirmed by experimental results, we suggest that the ratio
$\tau_{\max} / \SubText{\tau}{total}$ can be 
used as a metric to assess the controller robustness to additional
delays in the control loop. Different than the classical methods which
require one plot for each rational speed, $\SubText{\tau}{max}$ depends only on
the controller parameters. Hence, it does not change with operating
conditions making the proposed method much simpler to use than the
classical ones. Finally, it is worth noticing the method
proposed here can be easily extended to other types of speed
estimators (such as the back EMF methods) as well as to analyze the
stability of the set-point tracking response.

\ifCLASSOPTIONcaptionsoff
  \newpage
\fi

\section*{Acknowledgment}
This research work was supported in part by the Natural Sciences and
Engineering Research Council of Canada NSERC/Engage Program and the
National Research Council Canada NRC/IRAP Program.

%############################################################################
%\bibliographystyle{IEEEtr}
%%\bibliography{K:/ieee/publications/digital_kit/journals/p1_TSC_2009_emad_20091229/Bibliography/refs}
%\bibliographystyle{C:/applications/papers/IEEEtran/IEEEtranBST2/IEEEtran}
%\bibliography{C:/applications/papers/Bibliography/refs}
%\bibliography{C:/applications/papers/Bibliography/refs}
% \bibliography{/Users/egad/Documents/Papers/Bibliography/refs.bib}
%\bibliography{/Users/egad/Documents/Papers/Bibliography/refs.bib}
%\bibliography{/Users/egad/Documents/Papers/Bibliography/refs.bib}
\bibliographystyle{IEEEtran.bst}

%\begin{biography}[{\includegraphics[width=1in,height=1.25in,clip,keepaspectratio]{mshell}}]{Michael Shell}


%\vfill


\end{document}